\documentclass[final,3p,times,fleqn,twocolumn]{elsarticle} %this for viewing in two column format
\usepackage{graphics}
\usepackage{graphicx}
\usepackage{epsfig}

\usepackage{amssymb}
\journal{Optics Communications}

\begin{document}

\begin{frontmatter}


% use the tnoteref command within \title for footnotes;
% use the tnotetext command for theassociated footnote;
% use the fnref command within \author or \address for footnotes;
% use the fntext command for theassociated footnote;
% use the corref command within \author for corresponding author footnotes;
% use the cortext command for theassociated footnote;
% \author{Name\corref{cor1}\fnref{label2}}
% \address{Address\fnref{label3}}

\title{Time-resolved UV-IR pump-stimulated emission pump spectroscopy to probe collisional relaxation of the $8p\,^2P_{3/2}$ state of Cs I}

\author[Miami] {Mohammed Salahuddin}
\author[Miami] {Phill Arndt}
\author[Miami] {Jacob McFarland}
\author[Miami]{S.~Bur\c{c}in~Bayram \corref{cor1}}\ead{bayramsb@miamioh.edu}
\address[Miami]{Department of Physics, Miami University, 500 E. Spring Street, Oxford, OH 45056, USA}
%\address[label2]{}
%\address{}
\cortext[cor1]{Corresponding author}

\begin{abstract}
We describe and use a time-resolved pump-stimulated emission pump spectroscopic technique to measure collisional relaxation in a high-lying energy level of atomic cesium. Aligned $8p\,^2P_{3/2}$ cesium atoms were produced by a pump laser. A second laser, the stimulated emission pump, promoted the population exclusively to the $5d\,^2D_{5/2}$ level. The intensity of the $5d\,^2D_{5/2}\rightarrow6s\,^2S_{1/2}$ cascade fluorescence at 852.12 nm was monitored. The linear polarization dependence of the $6s\,^2S_{1/2}\rightarrow8p\,^2P_{3/2}\rightarrow5d\,^2S_{5/2}$ transition was measured in the presence of argon gas at various pressures. From the measurement, we obtained the disalignment cross section value for the $8p\,^2P_{3/2}$ level due to collisions with ground-level argon atoms.
\end{abstract}
\begin{keyword}
spectroscopy \sep atom-atom collisions \sep polarization
\PACS 32.00.00 \sep 32.90.+a \sep 32.50.+d
% MSC codes here, in the form: \MSC code \sep code
% or \MSC[2008] code \sep code (2000 is the default)
\end{keyword}

\end{frontmatter}

\section{\label{sec:intro}Introduction}
\setlength{\mathindent}{0pt}  %for flashing to the left for all equations
%\noindent
Collisional depolarization cross sections between
an alkali-metal and rare gas atom play important role in understanding the global
interactions between the two species. Studies of excited atom collisions with other neutral
atoms and molecules are key to understanding energy transfer processes,
anisotropic cross sections, and the accurate description of
the intermediate molecular properties of the colliding species~\cite{Baylis,Cook93,Lasell97,Wong98,Zhao07,Lin07,Bayram12,Bayram12-2}.
Collisional depolarization of the first excited $P_{3/2}$ state of cesium has been
experimentally studied using a pump-probe technique~\cite{Bayram2009} and
by using incoherent light sources as optical excitation~\cite{Guiry76,Fricke67}. The depolarization of the second excited $P_{3/2}$ states of Cs I and Rb I in collisions with rare gas atoms have been studied using a level-crossing technique~\cite{Lukaszewski83}. To our knowledge, there are no experimental observations of the disalignment cross section
for the third excited state ($8p\,^2P_{3/2}$) of cesium. We anticipate that this is due to the relatively high energy (3.198 eV) of the 8p level with respect to the first ionization limit ($\sim$ 3.89 eV) that makes the typical pump-probe (stepwise scheme) experimentally difficult.

Our experimental technique is based on a $\Lambda$-type double-resonance transition. In the $\Lambda$-type scheme the probe laser resonantly induces downward transitions from the upper $8p\,^2P_{3/2}$ level of the pumped transition to the $5d\,^2D_{5/2}$ level. This process is called stimulated emission pumping (SEP). The advantages of the PUMP-SEP scheme are that it can be used to probe either high-lying energy levels of atoms which are difficult to reach using stepwise excitation when the probed level is close to the ionization level, or high-lying electronic vibrational-rotational levels of molecules close to the dissociation or ionization limit of the electronic state under study. Thus, in this work, we have used a nanosecond PUMP-SEP technique which is particularly well suited for the study of the dynamical properties of atomic and molecular systems, e.g. Rydberg energy levels which are relatively inaccessible using a typical stepwise scheme.

In this work, we have measured the linear polarization dependence of the $6s\,^2S_{1/2}\rightarrow8p\,^2P_{3/2}\rightarrow5d\,^2D_{5/2}$ transition as a function of argon gas pressure using the PUMP-SEP technique. From the measurement we extracted the alignment-dependent collisional cross section of the $8p\,^2P_{3/2}$ cesium in collision with argon atoms. Our result yields a direct measure of the importance of the linear polarization to the alignment-dependent inelastic process in alkali-noble-gas collisions and provides additional insight into the collisional relaxation for the higher-lying energy levels of alkali atoms. An extensive theoretical treatment of collisional depolarization of atomic fluorescence has been developed by Ref.~\cite{RebaneRebane72,Rebane72}. Comprehensive reviews on the interpretation of atomic alignment and linear polarization, and the influence of spin-orbit coupling to the alignment parameter are discussed by Andersen and co-workers~\cite{Andersen97,Andersen}. Theoretical expressions for the rate constants of the anisotropic
collisional relaxation of atomic polarization moments in terms of multipole moments are derived in Ref.~\cite{Petrashen93}.

\section{\label{sec:concept}Experimental approach to measure polarization spectra}

In this section we provide a brief overview of the experimental scheme and the arrangement that was used to perform the experiments. The cesium transitions involved in this experiment are illustrated by the partial energy level diagram in Fig. 1 and a schematic overview of the experimental apparatus is shown in Fig.~2. We use a nanosecond pulsed neodymium-doped yttrium aluminum garnet (Nd:YAG) laser which operates simultaneously at 532 nm and 355 nm with a pulse repetition rate of 20 Hz. This laser is used to drive two home-built grazing-incidence Littman-Metcalf cavity design dye laser oscillators. The output from the third harmonic generator (355 nm) is used to produce a UV pump laser at 387.92 nm to populate the $8p\,^2P_{3/2}$ level while the output of the second harmonic generator (532 nm) is used to produce the IR probe laser at 894.72 nm to drive $8p\,^2P_{3/2}\rightarrow5d\,^2D_{5/2}$ transition as probe. The dye laser oscillators operate in a single transverse mode and the output beam is highly linearly polarized through use of Glan-Thompson calcite prism polarizers having extinction ratios of better than 10$^{-5}$. Both dye lasers are equipped with dye circulating systems to maintain an average power of about 2 mW.
%FIGURE 1
\begin{figure}[ht]
\begin{center}
{$\scalebox{1.60}{\includegraphics*{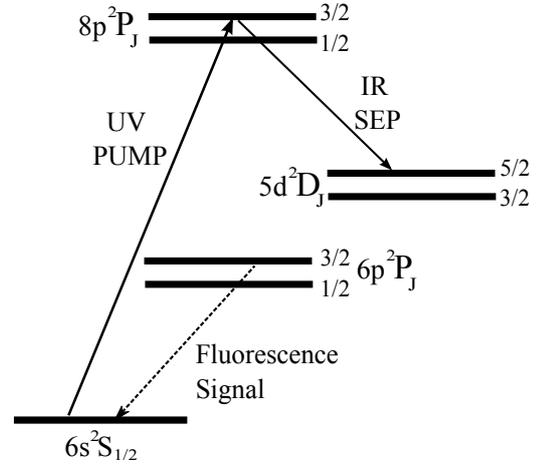}}$ }  %figure1.eps
\caption{Partial energy level diagram of Cs I showing the UV-IR PUMP-SEP transitions tuned to resonance at 387.92 nm and 894.72 nm, respectively. Population from the $5d\,^2D_{5/2}$ level excusively decays to the $6p\,^2P_{3/2}$ level. The dotted line shows the detected fluorescence signal observed at 852.12 nm.}
\end{center}
\label{fig1}
\end{figure}

A temperature-controlled liquid crystal variable retarder (LCR) is used to electronically vary the linear polarization direction of the probe laser to be parallel or perpendicular to that of the pump laser. Polarization switching of the LCR is achieved by applying the necessary voltage to the retarder via a computer-controlled liquid crystal digital interface. The beams of the pump and probe pulsed dye lasers are directed collinearly, but in opposite directions, into the interaction region of the cesium cell.  A resistively heated nonmagnetic cylindrical aluminum oven was used to generate the desired vapor pressure of atomic Cs in the cell. The oven, which houses the sealed Pyrex cell containing Cs vapor, was wrapped with an aluminum oxide blanket and an insulator to maintain the relative temperature to better than $\pm 0.01^{o}$C via a temperature controller. Cesium vapor cells with argon gas pressures ranging up to 133 mbar were prepared using an oil-free vacuum system in a 25.4 mm diameter by 50.8 mm length cell. The background pressure of the pure Cs cell is about $10^{-4}$ mbar.

After the satisfaction of the resonance in the $8p\,^2P_{3/2}$ level by UV PUMP laser the most of the population decays to the $5d\,^2D_{5/2}$ level by the IR SEP laser due to the stimulated emission~\cite{Domiaty94}. The intensities of the cascade fluorescence from the $5d\,^2D_{5/2}$ level to the ground $6s\,^2S_{1/2}$ level were recorded at 852.12 nm by using an infrared sensitive water-cooled photomultiplier tube (PMT) which was located at right angles to the propagation directions of the lasers. A combination of interference and color glass filters was used in front of the PMT in order to remove background light. All the cables used in the experiment were electrically shielded and the optical table was grounded in order to suppress electronic pick-up and noise on the observed signal.
%FIGURE 2
\begin{figure}[th]
\begin{center}
{$\scalebox{1.20}{\includegraphics*{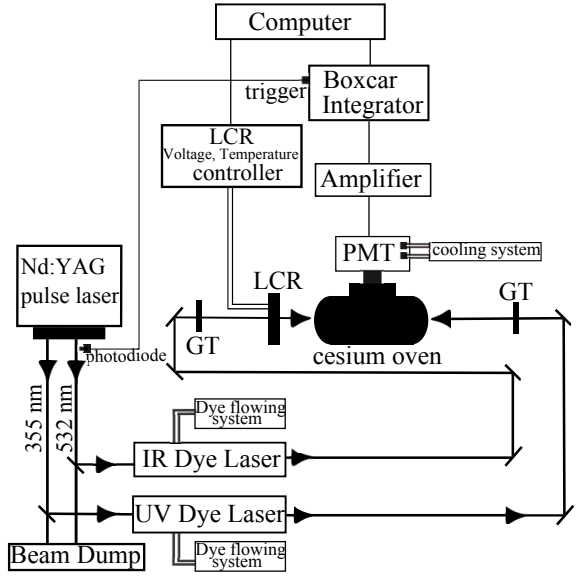}}$ }  %figure2.eps
\caption{A schematic view of the experimental apparatus.  In the figure GT stands for a Glan-Thompson polarizer, PMT for photomultiplier tube, and LCR refers to liquid crystal retarder.}
\end{center}
\label{fig2}
\end{figure}

The amplified output of the PMT signal was sent to a boxcar which was opened after a 1-ns delay following the laser pulses. The recorded signal collected for each state of laser polarization consisted of 100x10$^6$ data points
accumulated during 4 seconds. The boxcar operated in a 100 sample averaging mode, where the average single-shot level within the detection gate is digitized. Since the lifetimes of the
$5d\,^2D_{5/2}$ and $6p\,^2P_{3/2}$ levels are shorter than the lifetime of the $8p\,^2P_{3/2}$ level (305 ns~\cite{Rad85})
we distinguished the signal from the spontaneous emission decay of the
atoms from the $8p\,^2P_{3/2}$ level to the lower levels. Typical signal size was about 10$^{3}$ photons for each laser pulse. The digitized signals were stored on a computer using a LabVIEW program while monitoring the size of the signal within the gate-width in real time using a digital oscilloscope operating at 500 MHz with 2 GSa/s. Comparison of the signals, detected when the probe polarization angle is $\chi$ = 0 versus $\chi$ = $\pi$/2 allows definition of a linear polarization degree. A linear polarization degree is measured when intensities $I_{\parallel}(\chi=0)$ and $I_{\perp}(\chi = \pi/2)$ according to

\begin{equation}
P_L=\frac{I_{\parallel}(\chi = 0) - I_{\perp}(\chi = \pi /2)}{I_{\parallel}(\chi = 0) + I_{\perp}(\chi = \pi /2)}.
 \label{Eq1n}
 \end{equation}

Since an absolute intensity ratio of the signals is sensitive only to the
relative polarization directions of the lasers, any variations of the laser intensities with experimental factors such as absorbing medium density, fluorescence background, and sensitivity of
the gated boxcar integrator, collectively have negligible effect on the
intensity ratio.

For atoms with non-zero nuclear spin, the total angular
momentum of the atomic ensemble couples to the nuclear spin moment
to produce a new space-fixed total angular momentum $F$.
The total electronic angular momentum will then precess about $F$ so
that the initially prepared space-fixed frame is altered. This
precession is results from the hyperfine structure and affects the initially prepared alignment at $t$=0 by a pump laser. In the case of $^{133}$Cs, the coupling of nuclear spin $I=7/2$ with $J=3/2$ in the $8p\,^2P_{3/2}$ level introduces an
oscillating time dependence in the alignment. Therefore, the time evolution of the alignment under the influence of the hyperfine structure
can be evaluated. $P_L$, whose value depends on the hyperfine energy separations in the probed $8p\,^2P_{3/2}$ level, is the main quantity to be measured in the experiment and strongly depends on the time delay between the pump and probe laser pulses. This means that the excited level can be characterized by an overall population and the axially symmetric electronic alignment tensor component $\langle{A_o}\rangle$.  The quantum mechanical treatment of the detection of alignment and polarization of the emitted light in terms of alignment have been given in detail~\cite{Green82,Fano73}, and the applications to the polarization measurement using two-photon pulsed pump-probe excitation has been described in our earlier work~\cite{Bayram2009,Bayram06}. Thus, the linear polarization in terms of alignment is defined~\cite{Green82} as

\begin{equation}
P_L=\frac{-3 \langle{A_o}\rangle}{16- \langle{A_o}\rangle}.
 \label{Eq1}
 \end{equation}
The pump laser excitation on the
$6s\,^2S_{1/2}\rightarrow8p\,^2P_{3/2}$ transition creates an initial value of electronic alignment $\langle{A_o}\rangle= -4/5$.  In the absence of any perturbations this quantity evolves in time according to $\langle{A_o(t)}\rangle = \langle A_o(0)\rangle g^{(2)}(t)$, where the quantity $g^{(2)}(t)$ is the hyperfine depolarization coefficient and strongly depends on the temporal width of and arrival time of the pump and probe lasers into the interaction region of the atomic oven. Measurements of the overlap time of the pulses were made using two fast vacuum photodiodes with 500 ps rise time. At $t=0$, in the absence of any perturbations and systematic error, the theoretical value of polarization is 1/7 (14.29\%). The linear polarization degree evolves in time due to hyperfine structure in the excited level. We have measured the time evolution of polarization from $t=0$ to $t=150$ ns in our earlier work~\cite{Bayram2014}. At $t=1.5$ ns the linear polarization for the $6s\,^2S_{1/2}\rightarrow8p\,^2P_{3/2}\rightarrow5d\,^2D_{5/2}$ transition is obtained to be 12.4(5)\%. This measurement is in excellent agreement with theory~\cite{Bayram2014,Blum81}. Figure~3 illustrates typical fluorescence signal immediately after stimulated emission pump laser arrives into the interaction region. The sharp peak confirms depletion of the excited state population by the stimulated emission pump transition and this is because the cascade fluorescence from the $5d\,^2D_{5/2}$ level populates the $6p\,^2P_{3/2}$ level exclusively. The population of the $6p\,^2P_{3/2}$ level subsequently decays by spontaneous emission to the ground level. This appears as a sharp increase in the signal size within about 13 ns at FWHM, compared to the spontaneous emission which occurs over hundreds of nanoseconds. We opened a boxcar gatewidth at 30 ns to measure the full intensity of the cascade fluorescence from the stimulated emission signal only.
%FIGURE 3
\begin{figure}[ht] %fig.3
\centering
{$\scalebox{0.35}{\includegraphics*{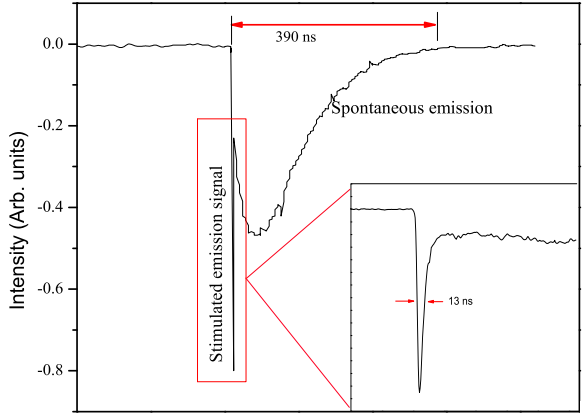}}$ }  %figure3.eps
\caption{Typical observed stimulated emission signal when the probe laser transfers the population from the $8p\,^2P_{3/2}$ to the $5d\,^2D_{5/2}$ level. Cascade fluorescence from the $5d\,^2D_{5/2}$ level populates the $6p\,^2P_{3/2}$ level exclusively. Inset shows the stimulated emission signal which causes substantial increase in observed cascade fluorescence when the probe laser beam follows the pump temporally.}
\label{signal}
\end{figure}

\section{Results and Discussions}
A linearly polarized pump laser selectively populates the Zeeman sublevels of the Cs $8p\,^2P_{3/2}$ level. After introducing argon atoms, the excited Cs and ground level Ar atoms collisionally mix the population among the Zeeman sublevels. Using rate equation analysis, population densities in terms of total population $N(t)$ and alignment $\langle{A_o(t)}\rangle$ were expressed in our earlier paper~\cite{Bayram2009}. Considering collisional mixing among the $M_J=\pm1/2,\pm3/2,\pm5/2$ sublevels, the measured signals for the $6s\,^2S_{1/2} \rightarrow 5d\,^2D_{5/2}$ transition can be written as
\begin{equation}
I_{\parallel}=\frac{1}{2}N(t)-\frac{1}{4}\langle{A_o(t)}\rangle,
 \label{integral1}
 \end{equation}
and
\begin{equation}
I_{\perp}=\frac{1}{2}N(t)+\frac{1}{16}\langle{A_o(t)}\rangle,
 \label{integral2}
 \end{equation}
where $N(t)=\frac{2}{\gamma}(1-e^{-\gamma t})$ and $\langle{A_o(t)}\rangle=\frac{-8}{5\gamma_a}(1-e^{-\gamma_a t})$. Substituting Eqs.~(\ref{integral1}) and (\ref{integral2}) into Eq.~(\ref{Eq1n}), a linear polarization can be readily expressed in terms of the depolarization cross section and the pressure of the buffer gas as
\begin{equation}
P_{L}=\frac{1}{3+4Z^{\prime}},
\label{polZ}
\end{equation}
where
\begin{equation}
Z^{\prime}=\frac{\gamma_a}{\gamma}~\frac{\left[1-\frac{1}{\gamma~T}(1-e^{-\gamma~T})\right]}
{\left[1-\frac{1}{\gamma_a~T}(1-e^{-\gamma_a~T})\right]}.
\label{Z}
\end{equation}
In Eq.~(\ref{Z}), $\gamma$ is the radiative decay rate, T is the overlap time of the pulses, and $\gamma_a$ is defined as $\gamma$+$\Gamma$ where $\Gamma$=$\rho_{Ar}k_2$ is the collisional rate. Here $k_{2}=\langle{\sigma_{2}v}\rangle$ is the disalignment rate coefficient, $\rho_{Ar}$ is the argon density which depends on the argon pressure and thermal energy constant, and $\sigma_{2}$ is the disalignment cross section. It is assumed that $\langle{\sigma_{2}v}\rangle$ may be factored since $\sigma_{2}$ typically depends weakly on the velocities of the colliding Cs-Ar atoms so that $k_2=\sigma_{2}\langle v\rangle$, where $\sigma_{2}$ is the disalignment cross section. Thus, we denoted
$\langle v\rangle$ as $\bar{v}_{CsAr}$ which is the average
speed of the colliding Cs-Ar atoms over the Maxwell-Boltzmann distribution of relative velocities at the
$80^{o}$C cesium cell temperature. The left side of Eq.~(\ref{polZ}) is the measured linear polarization value at various argon gas pressures. Measured linear polarization degree as a function of argon gas pressures ranging up to 133 mbar is illustrated in Fig.~4. The result of the best fit to the data from the weighted non-linear least-squares fitting program is shown as the solid line. From the measurement $\sigma_{2}$ was extracted.
%FIGURE 4
\begin{figure}[ht] %fig4
\centering
{$\scalebox{0.28}{\includegraphics*{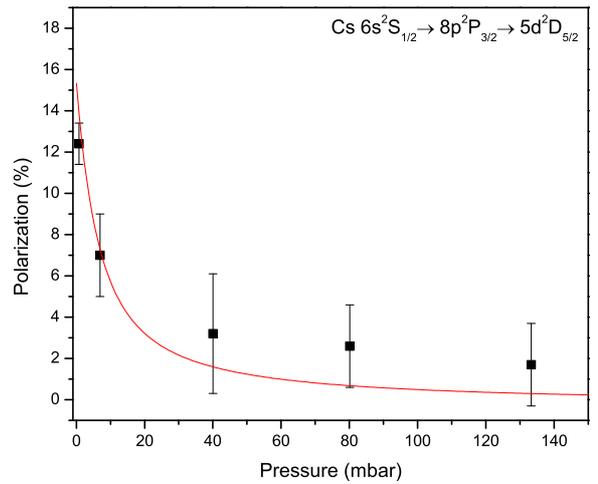}}$ }  %figure4.eps
\caption{Nonlinear least square fit of the measured polarization as a function of argon gas pressure. Vertical error bars represent one standard deviation.}
\label{fitting}
\end{figure}
\\

%\subsection{Discussion}
Collisional depolarization cross sections for the lowest few principal quantum numbers of the Cs I $np\,^2P_{3/2}$ series are summarized in Table I. Experimental results are obtained by a broadened level crossing technique~\cite{Lukaszewski83,Minemoto74} or by a pump-stimulated probe approach, as described in the present report.  Also listed are theoretical determinations of the cross-sections. We see in Table I that the cross section increases sharply for the $7p\,^2P_{3/2}$ level in comparison with the $6p\,^2P_{3/2}$; the further increase associated with the measurements made here on $8p\,^2P_{3/2}$ level is much more modest.  This is somewhat surprising given the typical sharp variations, with effective principal quantum number, of experimental observables associated with excited atomic levels.  However, previous observations of collisional process cross sections have revealed nonmonotonic behavior in some cases. For instance, Gallagher~\cite{Gall94} reported that the cross sections for the collisional angular momentum mixing rise approximately as $n^4$ for lower $n$, reach a peak, and then decreases for larger $n$. Other researchers have reported oscillatory cross sections in some cases and large oscillations in the dependence of linewidth on principal quantum number~\cite{Hugon79,Stoicheff80}. These results motivate us to explore disalignment cross section measurements for a wider range of principal quantum numbers; the technique used here is ideally suited for such measurements.
\begin{table} [ht]
\caption{\label{tab:depolarizationtab1}Collisional cross section
($\sigma_{2}$) of the $np\,^2P_{3/2}$ Cs atom in collision with Ar
buffer gas is listed. Here, PPS refers to pump-probe spectroscopy and BLC broadening of level crossing.}
\begin{tabular}{l@{}l@{\hspace{5mm}} l@{}l@{\hspace{5mm}} l@{}l@{\hspace{5mm}}l}
\hline\noalign{\smallskip}
\multicolumn{2}{l@{\hspace{11mm}}}{$n$} &
\multicolumn{2}{l@{\hspace{11mm}}}{$\sigma_{2}(\AA^2)$} &
\multicolumn{2}{l@{\hspace{11mm}}}{Technique} &
\multicolumn{1}{l}{Reference} \\
\hline\noalign{\smallskip}
6  &          & 186(58) &    & PPS &     &\cite{Bayram06} \\
   &          & 238~~~~~~~&    & Theory&    &\cite{Okunevich70}\\
7  &          & 730(24) &    & BLC &    &\cite{Lukaszewski83}\\
   &          & 610~~~~~~~&    & BLC &    &\cite{Minemoto74}\\
   &          & 557~~~~~~~&    & Theory&    &\cite{Okunevich70}\\
8  &          & 895(57) &    & PPS &     &This work\\
   \hline\noalign{\smallskip}
\end{tabular}
\end{table}
Our result is consistent with the lower $n$ level values. However, future experiments on greater $n$ values could reveal any systematic structure on the $n$ dependence.
\section{\label{sec:results}Conclusions}
We have introduced a time-resolved double-resonance UV-IR nanosecond PUMP-SEP technique to provide experimental data on the depolarization cross section of the cesium $8p\,^2P_{3/2}$ level from the measurement of polarization degree of the $6s\,^2S_{1/2}\rightarrow8p\,^2P_{3/2}\rightarrow5d\,^2S_{5/2}$ transition. From the measurement at various gas pressures we extracted the collisional disalignment cross section using nonlinear least square fit to the data and obtained 785(57)$\AA^2$. This work complements and substantiates our earlier studies of the collisional disalignmnet cross section of the excited states of cesium at lower $n$ values. We anticipate that this technique will be applicable to measure collisional energy transfer in higher energy levels of many atoms and in diatomic molecules.

\section*{Acknowledgement}
Financial support from the National Science Foundation (Grant No. NSF-PHY-1309571) is gratefully acknowledged. The authors would like to thank Greg Reese, from Research Computing Support, for providing the MATLAB programming code to do the weighted nonlinear least square fit to the data. We would like to thank Professor Mark Havey of Old Dominion University for valuable discussions.

%\section*{References}  %here * means do not use Roman numbers or anything.
%\bibliography{depolarization}% Produces the bibliography via BibTeX. Go to *.bbl file to paste it here below for submission process.
%\bibliography{collision}
%***********************************************************************************************

%**********************************************************************************************
%CAPTIONS

\end{document}